\documentclass{article}
\usepackage{graphicx} %
\usepackage{subfig}
\usepackage{amsmath,amsfonts,amssymb,amsthm,bm,mathtools,graphicx,subfig,color,tcolorbox,algorithm,algorithmic}
\usepackage{caption,fancyhdr,titlesec,indentfirst,booktabs,verbatim,hyperref,cases,titletoc,multirow,wasysym,empheq,setspace,chngcntr,authblk}
\usepackage[authoryear,sort]{natbib}

\title{\Large{Li, Li, and Dai's Contribution to the Discussion of “Estimating Means of Bounded Random Variables by Betting” by Waudby-Smith and Aaditya Ramdas}}
\author[1]{Jiayi Li\thanks{These authors contributed equally to this work.}}
\author[1]{Yuantong Li$^{*}$}
\author[1]{Xiaowu Dai\thanks{Corresponding Email: dai@stat.ucla.edu.}}
\affil[1]{Department of Statistics, University of California, Los Angeles}
\date{}

\usepackage{xcolor}         %
\usepackage{hyperref}       %
\usepackage{caption}

\begin{document}

\maketitle

\vspace{-0.1in}

\noindent
We congratulate Waudby-Smith and Ramdas for their interesting paper \cite{waudbysmith2022estimating} in generating confidence intervals and time-uniform confidence sequences for mean estimation with bounded observations.  Their methodology utilizes composite nonnegative martingales and establishes a connection to game-theoretic probability.
Our comments will focus on numerical comparisons with alternative methods. The corresponding code is available at \url{https://github.com/Likelyt/Estimate-mean-with-betting}.

\section{Methods}
\label{sec:methods}
\noindent
Consider a sequence of random variables $(X_{t})_{t=1}^{\infty}$, drawn from a distribution $P \in \mathcal{P}^{\mu}$, where $\mathcal{P}^{\mu}$ represents the set of all distributions on $[0,1]^\infty$. We assume $\mathbb{E}_{P}[X_{t}|X_1,\ldots,X_{t-1}] = \mu$ for some unknown $\mu\in[0,1]$. The objective is to construct a time-uniform confidence sequence $(C_t)_{t=1}^\infty$ that satisfies the condition
\begin{equation}
\label{eqn:tucs}
    \sup_{P\in \mathcal{P}^{\mu}}P(\exists t\geq 1:\mu\not\in C_t)\leq \alpha.
\end{equation}
The confidence sequence guarantees that, for any fixed $n$, $C_n$ is a $(1-\alpha)$-confidence interval for $\mu$.
We review three methods for generating such confidence sequences.
\begin{itemize}
\item[(a)] The \emph{predictable plug-in empirical Bernstein} (Pr-EB) method, discussed in Section 3 of \cite{waudbysmith2022estimating}, combines the Robbins' method of mixture \cite{robbins1970statistical} with exponential supermartingales.

\item[(b)] The \emph{hedged capital process} (Betting) method, introduced in Section 4 of \cite{waudbysmith2022estimating}, is a \emph{novel} approach that enjoys the interpretation of wealth accumulation in a game and has connections to the game-theoretic probability \cite{shafer2005probability}.

\item[(c)] The \emph{Bootstrap} resampling method is implemented using the R package \textsc{boot} \cite{bootpackage1}. With $B=200$ bootstrap replicates and $L=10$ batches, we calculate a separate confidence interval for data sequence $2^{l}\leq t<2^{l+1}$ within each batch $l=1,\ldots, L$. The confidence intervals are constructed using the $\frac{\alpha}{2L}$- and $(1-\frac{\alpha}{2L})$-quantiles of the bootstrap means.

\end{itemize}

\section{Numerical Studies}

\paragraph{Synthetic Data Example}
We sequentially generate data from $\text{Beta}(10,30)$ distribution for $1\leq t\leq 10^4$, %
and construct a $95\%$ confidence sequence $C_t$ in Eq.~\eqref{eqn:tucs} using methods (a)-(c) in Section \ref{sec:methods}.  %

\begin{figure}[!ht]
    \centering
    \subfloat[Confidence sequence]{%
        \includegraphics[width=0.5\linewidth]{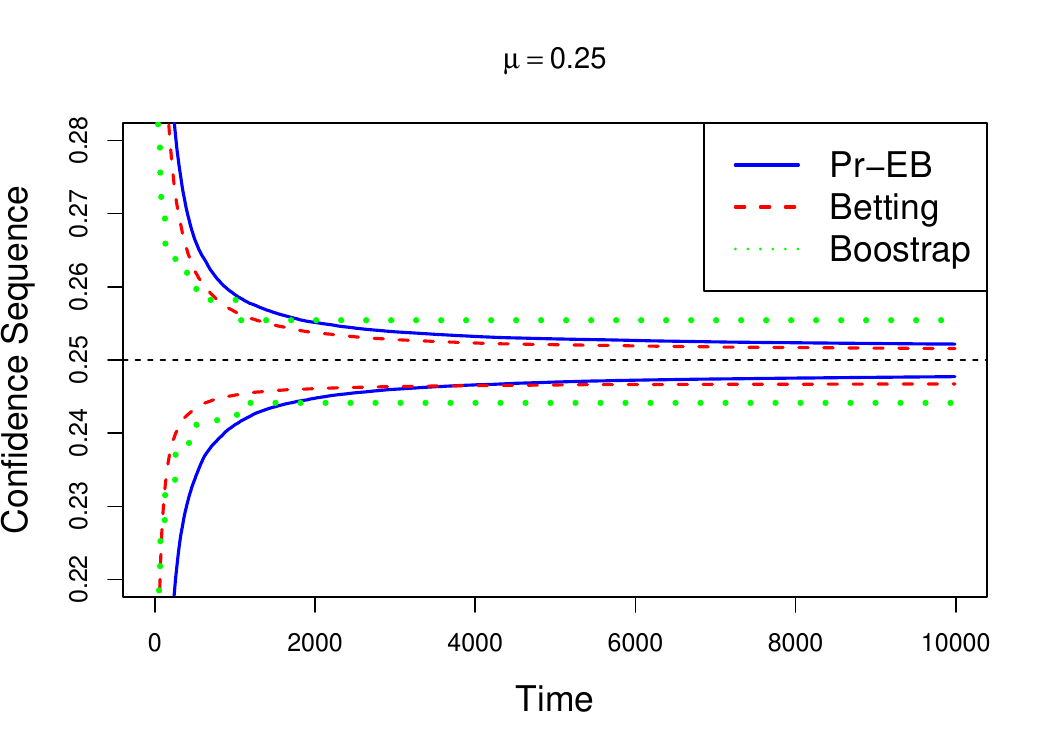}%
        }%
    \subfloat[Coverage length]{%
        \includegraphics[width=0.5\linewidth]{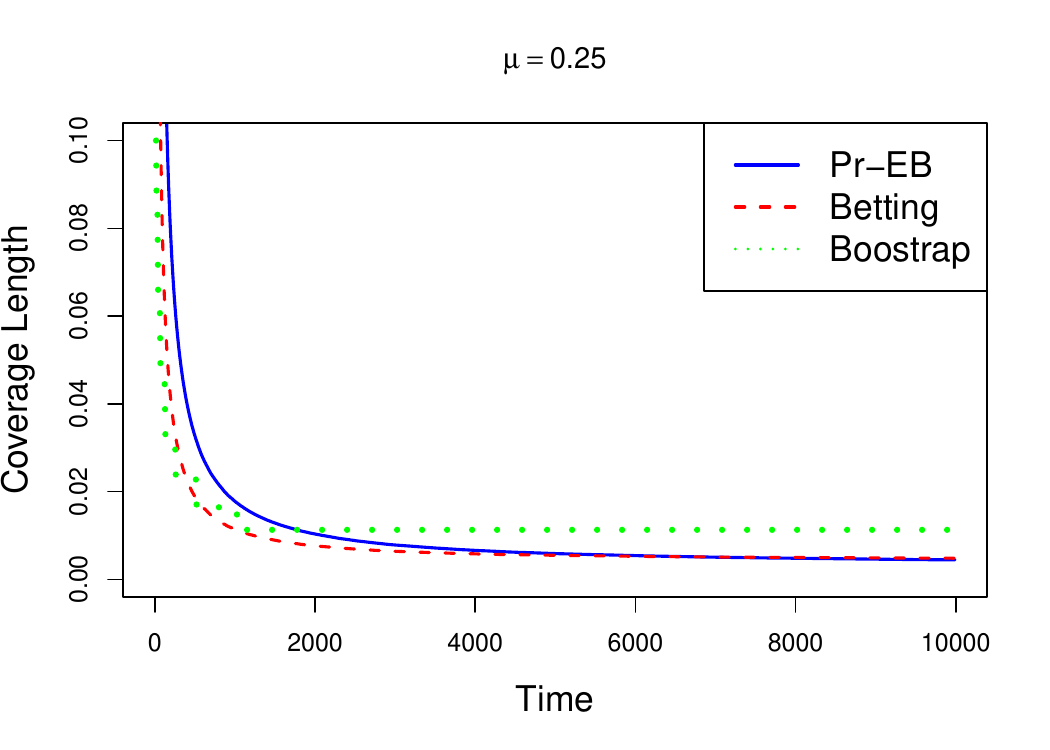}%
        }%
    \caption{Comparisons based on the synthetic data of Beta distribution.%
    }
    \label{D1}
\end{figure}

Figure~\ref{D1} shows that the Betting method outperforms the Bootstrap method with a higher lower bound in the confidence sequence and consistently tighter intervals for $t\geq 1500$. This aligns with the intuition that the Bootstrap method results in wider intervals due to dividing the confidence budget among data subsets when dealing with a large number of sequential data points.
Moreover, the Betting method outperforms the Pr-EB method with consistently tighter confidence sequences for $t\leq 3000$. For $t\geq 3000$, both methods yield comparable coverage lengths. The Betting method's sequence shows a slight shift towards smaller values, reflecting the left-skewed ground truth distribution, while the Pr-EB method produces a symmetric interval around the estimated mean.

\paragraph{Real Data Example}
We analyze a batting dataset of 18 Major League players from the 1970 season, available in \cite{efron2021computer} or the R package \textsc{EfronMorris}. The goal is to construct a $95\%$ confidence interval for the true batting level of each player based on their first 45 at-bats. The following results are obtained from 100 data replications.

\begin{figure}[!ht]
    \centering
    \subfloat[Confidence interval]{%
        \includegraphics[width=0.5\linewidth]{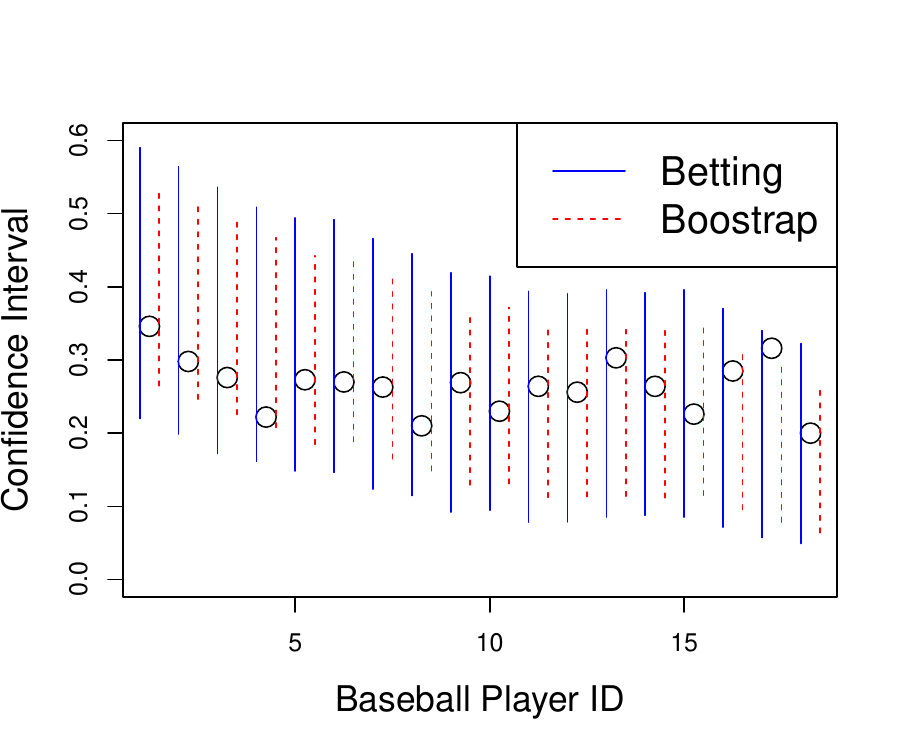}%
        }%
    \subfloat[Coverage Probability]{%
        \includegraphics[width=0.5\linewidth]{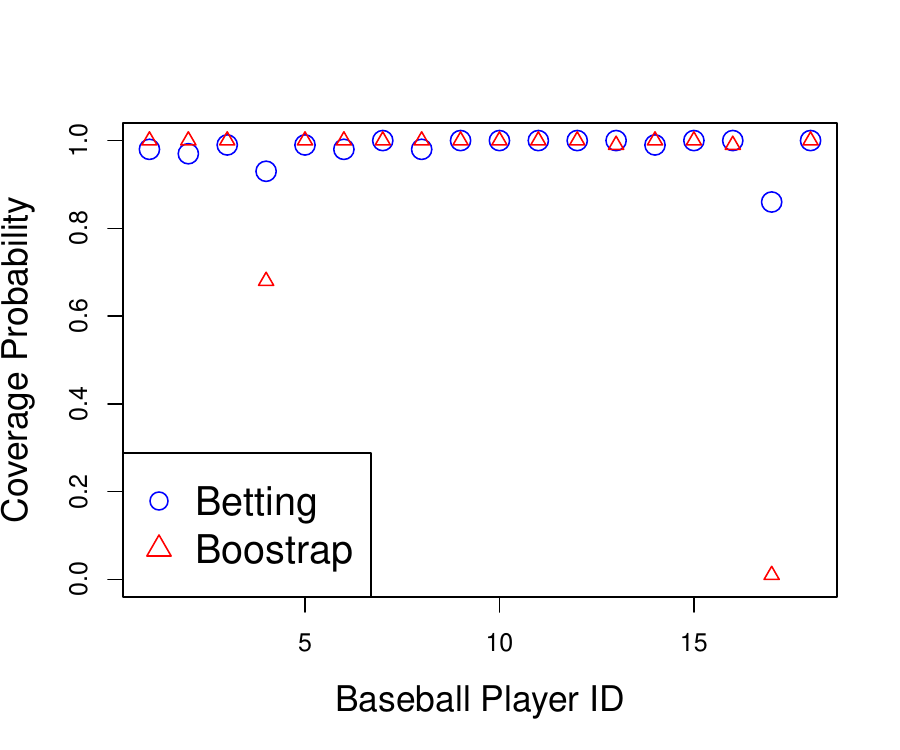}%
        }%
    \caption{Comparisons based on the baseball batting data from \cite{efron2021computer}.
    }
    \label{D2}
\end{figure}

Figure \ref{D2}(a) shows the average confidence intervals, where the black circle represents the true batting level of each player. %
Note that while the Bootstrap method fails to cover the true batting level of player 17, the Betting method successfully includes it. 
Figure \ref{D2}(b) presents the coverage probability of both methods. The Betting method achieves higher coverage probability for the true batting levels of players 4 and 17 compared to the Bootstrap method. These results indicate that the Betting method outperforms the Bootstrap method in constructing accurate confidence intervals for this baseball data.

\section{Extensions}
\noindent
The work in \cite{waudbysmith2022estimating} may inspire several directions for future research. One such direction is the generalization of the Betting method to handle multi-dimensional observations. Currently, the Betting method relies on the assumption that the underlying capital process is a martingale. However, when estimating the mean of multidimensional data, applying Ville's inequality \cite{NEURIPS2020_e96c7de8} for confident rejection of the null hypothesis becomes more challenging. Defining the hedged capital process in a vector space introduces complexities when simultaneously estimating multiple attributes.
Another direction is the extension of the Betting method to online decision-making scenarios, such as dynamic treatment, online recommendation, and dynamic pricing. These tasks frequently involve constructing confidence sequences to determine optimal actions. The application of the Betting method in these contexts seems to be interesting.

\bibliographystyle{apalike}
\bibliography{references}

\end{document}